\documentclass[aps,prd,superscriptaddress,long,notitlepage,balancelastpage,nofootinbib,floatfix,twocolumn]{revtex4-1}
\pdfoutput=1
\usepackage{amsmath,mathtools,amssymb,amsthm,amsxtra,overpic,bbm,epsfig,subfigure,url,bm}
\usepackage{hyperref}
\usepackage{mathrsfs}
\usepackage{color,xcolor}
\usepackage{comment}
\usepackage{float}
\usepackage{enumitem}
\usepackage{slashed}
\usepackage{multirow}
\usepackage[normalem]{ulem}
\definecolor{nicered}{rgb}{0.5,0.,0.}
\definecolor{nicegreen}{rgb}{0.,0.5,0.}
\definecolor{niceblue}{rgb}{0.,0.,0.5}
\hypersetup{
	colorlinks=true,
	linkcolor=black,
	filecolor=nicegreen,      
	urlcolor=niceblue,
	citecolor=nicered,
}

\makeatletter
\newcommand*{\balancecolsandclearpage}{%
	\close@column@grid
	\cleardoublepage
	\twocolumngrid
}
\makeatother
\begin{document}

\title{\vspace{1cm} \Large 
Resonances of Supernova Neutrinos \\ in Twisting  Magnetic Fields 
}

\author{\bf Sudip Jana}
\email[E-mail:]{sudip.jana@mpi-hd.mpg.de}
\affiliation{Max-Planck-Institut f{\"u}r Kernphysik, Saupfercheckweg 1, 69117 Heidelberg, Germany}

\author{\bf Yago Porto}
\email[E-mail:]{yporto@ifi.unicamp.br}
\affiliation{Instituto de F{\'i}sica Gleb Wataghin - UNICAMP, 13083-859, \\ Campinas, S\~ao Paulo, Brazil}

\begin{abstract}
We investigate the effect of resonant spin conversion of the neutrinos induced by the geometrical phase in a twisting magnetic field. We find that the geometrical phase originating from the rotation of the transverse magnetic field along the neutrino trajectory can trigger a resonant spin conversion of Dirac neutrinos inside the supernova, even if there were no such transitions in the fixed-direction field case. We have shown that even though resonant spin conversion is too weak to affect solar neutrinos, it could have a remarkable consequence on supernova neutronization bursts where very intense magnetic fields are quite likely. We demonstrate how the flavor composition at Earth can be used as a probe to establish the presence of non-negligible magnetic moments, potentially down to $10^{-15}~\mu_B$ in upcoming neutrino experiments like the Deep Underground Neutrino Experiment (DUNE), and the Hyper-Kamiokande (HK). Possible implications are analyzed.

\noindent 
\end{abstract}
\maketitle
\textbf{\emph{Introduction}.--} Wolfgang Pauli's 1930 letter \cite{Pauli:1930pc} not only postulates the neutrino's existence as an explanation for the apparent non-conservation of energy in radioactive decay but also suggests that these elusive particles possess a mass, along with non-zero magnetic moments. Later, in 1954, Cowan \textit{et al.} set the first limits on neutrino magnetic moments \cite{Cowan:1954pq}, even before neutrinos were discovered and Bernstein \textit{et al.} did a survey \cite{Bernstein:1963qh} of the experimental information on neutrino electromagnetic properties. In the late 80s and early 90s, the study of neutrino magnetic moments became a popular topic in addressing the solar neutrino problem \cite{Cisneros:1970nq, Okun:1986na, Lim:1987tk, Akhmedov:1988uk}. In recent decades, several experiments have discovered neutrino oscillations, which conclusively demonstrated that neutrinos have masses and mixing, indicating the need for physics beyond the standard model (BSM). In such BSM theories, neutrinos establish interactions with photons via quantum loop correction, even though neutrinos are immune to electromagnetic interaction in the Standard Model (SM)\footnote{In SM, the chiral symmetry obeyed by massless neutrinos demands $\mu_\nu=0$. In the minimal $\text{SM}+\nu_R$ scenario, $\mu_\nu$ is expected to be:    $\mu_\nu = \frac{ e G_F m_\nu}{8 \sqrt{2}\pi^2} = 3 \times 10^{-20} \mu_B\, \left(\frac{m_\nu}{0.1~{\rm eV}}\right)~$\cite{Fujikawa:1980yx}.
}. There are now hundreds of potential neutrino mass models. Yet, not all models qualify to conceive large magnetic moments without upsetting neutrino masses (see Ref.~\cite{Babu:2020ivd} and references therein). 
Neutrinos with large magnetic moments can significantly impact searches at neutrino scattering experiments \cite{Beda:2012zz, Borexino:2017fbd, CONUS:2022qbb} and dark matter direct detection experiments \cite{Aprile:2020tmw, XENON:2022ltv, LZ:2022ufs}, astrophysical neutrino signals \cite{Jana:2022tsa, Akhmedov:2022txm, Kopp:2022cug, Ando:2002sk, Ahriche:2003wt,  Akhmedov:2003fu}, stellar cooling \cite{Viaux:2013lha, Viaux:2013hca,  Capozzi:2020cbu}, cosmological imprints \cite{Vassh:2015yza, Li:2022dkc, Grohs:2023xwa}  and charged lepton's magnetic moment \cite{Babu:2021jnu} (for a review, see Ref.~\cite{Giunti:2014ixa}). The presence of large transverse magnetic fields within the sun, supernovae, neutron stars, or other astrophysical objects can result in efficient spin-precession \cite{Cisneros:1970nq, Okun:1986na} or resonant spin-flavor precession \cite{Lim:1987tk, Akhmedov:1988uk} of neutrinos. 

The majority of the literature (see Ref.~\cite{Jana:2022tsa, Akhmedov:2022txm, Kopp:2022cug, Ando:2002sk, Ahriche:2003wt,  Akhmedov:2003fu} and references therein), however, assumes that the direction of the transverse magnetic field is fixed. Nevertheless, this is not always the case. In such scenarios, neutrinos traveling from the core of the supernova outwards and crossing such field configurations would encounter a transverse magnetic field whose direction changes continuously throughout their trajectory \cite{Bugli:2019rax}. It will introduce a new geometrical phase\footnote{The  Berry phase \cite{Berry:1984jv} arises from the time-dependency of parameters entering the Hamiltonian of a quantum system.} governed by the magnetic field rotation angle $\phi$ in addition to the usual dynamical phase, determined by the energy splitting of the neutrino eigenstates. In the early 1990s, although Vidal \textit{et al.} and Aneziris \textit{et al.} discussed the effect of such phases in the context of solar neutrino problem \cite{Vidal:1990fr, Aneziris:1990my}; Smirnov was the first to correctly recognize \cite{Smirnov:1991ia} the resonant structure of neutrino spin-precession due to the solar magnetic field's geometrical phase, which was later explored by other authors \cite{Smirnov:1991ia, Akhmedov:1991vj, Akhmedov:1993ta, Balantekin:1993ys}.
However, we find that the impact of the geometrical phase on neutrino precession in the Sun is negligible, given a magnetic moment less than $\sim 10^{-11}~\mu_B$, which has been ruled out by current laboratory-based experiments \cite{XENON:2022ltv, LZ:2022ufs, CONUS:2022qbb}. One needs a perhaps unrealistically large magnetic field and magnetic moment combination for an emphatic effect. Here, we have shown that it could have a remarkable consequence on supernova neutronization bursts where very intense magnetic fields are quite likely. We analyze the neutrino spectra from the neutronization burst phase and demonstrate how its time-variation can be used as a probe to establish the presence of non-negligible magnetic moments, potentially down to $10^{-15}~\mu_B$  in forthcoming neutrino experiments like DUNE \cite{DUNE:2020zfm}, and HK \cite{Hyper-Kamiokande:2021frf}. Considering a realistic setup, it will be extremely difficult to probe $\mu_\nu \lesssim 10^{-12}~\mu_B$ in laboratory-based experiments based on neutrino-electron scattering. Moreover, it has been argued that Dirac neutrino magnetic moments over $10^{-15}\mu_B$  would not be natural \cite{Bell:2005kz} because they would produce unacceptable neutrino masses at larger loops. Thus, the results presented here are in the borderline of the conceivable region for $\mu_\nu$.

\begin{figure*}[t!]
\centering
  \includegraphics[width=0.7\textwidth]{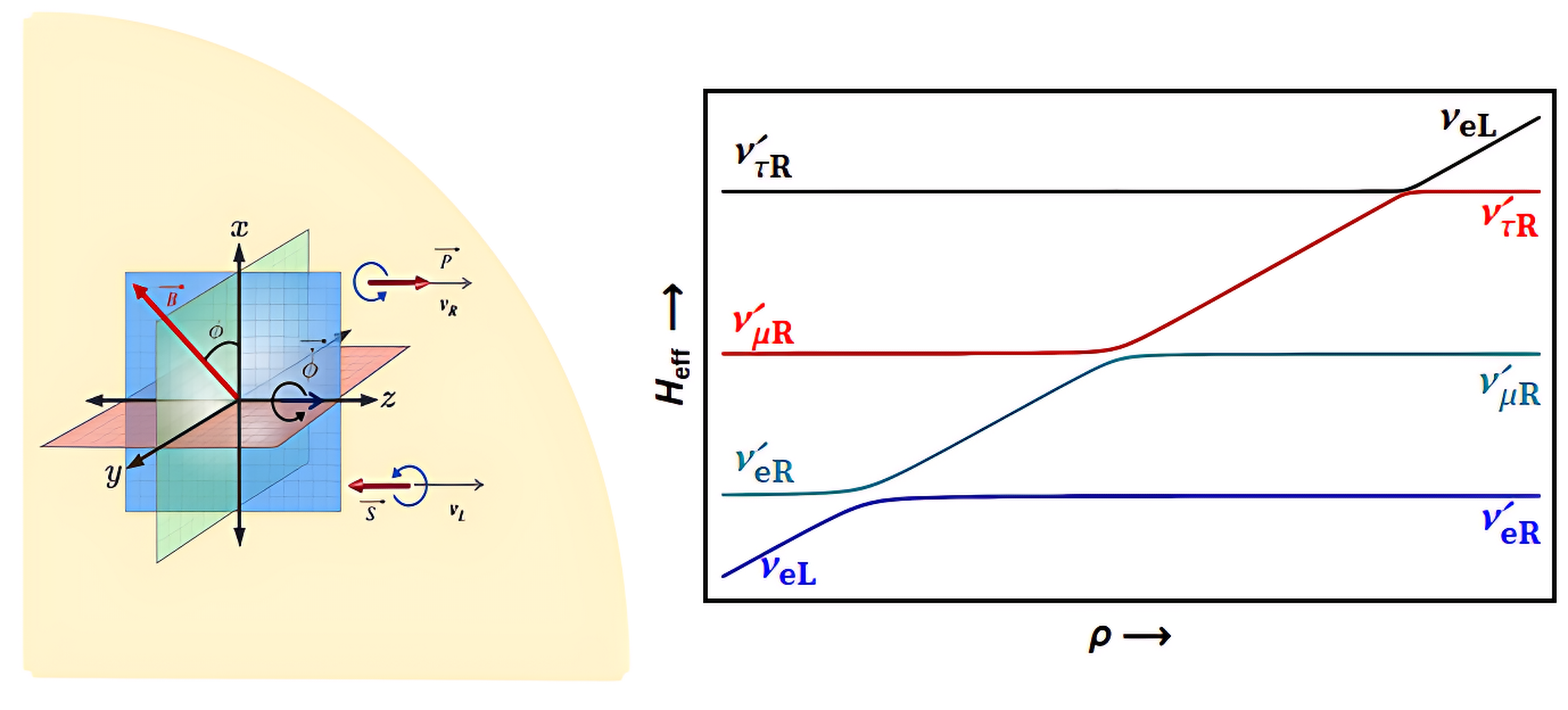}
  \caption{Left: schematic representation of the rotating frame where z-axis denotes the direction of the neutrino momentum, $\dot{\phi}$ is the velocity of $B$ field-rotation. The spins of $\nu_{L,R}$ are shown by thick red arrows. Right: neutrino energy levels
  in the resonance region for Normal Ordering (NO) and $\dot{\phi}<0$. See text for details.}
\label{energy-levels}
\end{figure*}

\textbf{\emph{Evolution of neutrino system in twisting magnetic field}.--} Let us consider a system of left-handed neutrinos $\nu_L=(\nu_{eL},\nu_{\mu L}, \nu_{\tau L})$, and right-handed counterparts $\nu_R=(\nu_{eR},\nu_{\mu R}, \nu_{\tau R})$, with magnetic moment $\mu$ evolving in matter and a transverse magnetic field.  If the magnetic field rotates along the neutrino path in the transverse plane, denoted by $\textbf{\textsl{B}}= B_x +i B_y = B e^{i\phi}$, where $\phi(r)$ is the angle of rotation, the resulting evolution equation can be expressed as:
\begin{equation} \label{eq:evolution}
    i \frac{d}{d r}\left[\begin{array}{c}
\nu_L \\
\nu_R
\end{array}\right]=\left[\begin{array}{cc}
H_L + (\dot{\phi}/2) I &  \mu B(r) \\
\mu^\dagger B(r)  & H_R - (\dot{\phi}/2) I
\end{array}\right]\left[\begin{array}{c}
\nu_L \\
\nu_R
\end{array}\right],
\end{equation}
where $r$ is the radial coordinate, $I$ is the identity matrix and $\mu=\text{diag}\{ \mu_\nu, \mu_\nu, \mu_\nu \}$ is the matrix of Dirac magnetic moments. Eq.~(\ref{eq:evolution}) is expressed  in a frame \cite{Akhmedov:1991vj} rotating with the magnetic field; see Fig.~\ref{energy-levels}.
$H_L$ represents  the Hamiltonian for $\nu_L$ propagating in matter, given by $ H_L= \frac{1}{2E} U \Delta U^{\dagger} + V,$
where $U$ is the PMNS matrix, $\Delta=\text{diag}\{0,\Delta m^2_{21},\Delta m^2_{31} \}$ and $V=\text{diag}\{ V_e, V_\mu, V_\tau \}$  is the matter potential. Assuming charge neutrality, $V_e=\sqrt{2}G_F (n_e-0.5 n_n)$ and $V_\mu=V_\tau=-0.5\sqrt{2}G_F n_n$, where $n_e, n_p, n_n$ are the number densities of electron, proton and neutrons respectively and $Y_e = n_e/(n_p+n_n)$ is the electron-fraction. $H_R$ is the Hamiltonian for $\nu_R$, which does not experience matter interactions, and $H_R= \frac{1}{2E} U \Delta U^{\dagger}$. For antineutrinos, $\Bar{\nu}_L$ are the ones that do not interact with matter. Furthermore, matter potentials for antineutrinos have the opposite sign: $\bar{V}_{e}=-V_e$ and $\bar{V}_{\mu, \tau}=-V_{\mu,\tau}$.

Now consider a neutrino system propagating in a background of the non-uniform matter and a rotating transverse magnetic field. As $V$ changes, strong resonant spin-flip conversion, $\nu_{L} \leftrightarrow \nu_{R}$, can occur. In the two-state approximation, the resonance condition for the $\nu_{\alpha L} \leftrightarrow \nu_{\alpha R}$ conversion  can be expressed as \cite{Akhmedov:1991vj}
\begin{equation} \label{resonance}
    V_\alpha+\dot{\phi}=0.
\end{equation} 
For antineutrinos, $\bar{\nu}_{\alpha L} \leftrightarrow \bar{\nu}_{\alpha R}$  resonance  condition is $\Bar{V}_\alpha - \dot{\phi} = 0$  and occurs at the same location. The dynamics  of spin-flip transition in the resonance region is governed by the adiabaticity coefficient $\gamma_\alpha$. Under two-state approximation, it can  be expressed as \cite{Akhmedov:1991vj}  
\begin{equation} \label{gamma}
    \gamma_\alpha = \frac{2(2 \mu_\nu B)^2}{|\dot{V}_\alpha+\ddot{\phi}|}.
\end{equation}
For our analysis, we study the three-flavor evolution (six states), described in Eq.~(\ref{eq:evolution}).
In such a scenario, coupled resonances (in which one resonance interferes with others owing to closeness) will exist between $\nu_{\alpha L}$ and all $\nu_R$ states since the right-handed (RH) states are linked among themselves as a result of mixing. The simplified neutrino energy levels in the resonance zone are depicted in Fig~\ref{energy-levels}. At high densities (to the right), $\nu_{eL}$ is heavier than all RH states since $V_e+\dot{\phi}(>0)$ is very large, but the converse occurs at low densities (to the left) where $V_e+\dot{\phi}<0$. The primed states in Fig.~\ref{energy-levels} are the eigenstates of $H_R$ and, to a decent approximation, of the whole system.

\textbf{\emph{Supernova environment}.--}
We focus on neutrinos released during the neutronization-burst phase, which occurs right after the core bounces and lasts for a few-tens of milliseconds. During this phase, the $\nu_e$-flux is dominant over other flavors in most of the energy-spectrum \cite{deGouvea:2019goq,Jana:2022tsa}. Moreover,  collective neutrino oscillations, a significant source of complication to the flavor evolution, are expected to be suppressed during this stage \cite{deGouvea:2012hg,Mirizzi:2015eza}. The predicted neutrino-fluxes during this phase have only about $\mathcal{O}(10 \%)$ uncertainty \cite{Serpico:2011ir,Wallace:2015xma,OConnor:2018sti,Kachelriess:2004ds}. As a result, we anticipate that the estimates of the SN neutrino flavor content during this period are more robust.

\begin{figure}[ht!]
\centering
  \includegraphics[width=0.45\textwidth]{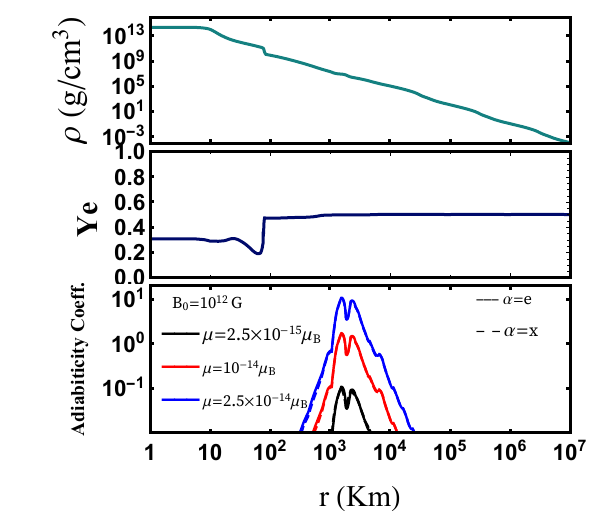}
  \caption{Matter-density $\rho$ (upper panel), electron number fraction $Y_e$ (middle panel) and adiabaticity coefficient $\gamma_\alpha$ (lower panel) as a function of the radial coordinate $r$. 
  }
\label{matter-profile}
\end{figure}

In this work, we explore the discovery potential of the Dirac magnetic moments of neutrinos coming from SNe with magnetic field strength $10^{10} - 10^{12}$ G in the iron-core. Such magnetic fields are commonly associated with the formation of magnetar-like field structures in SN remnants \cite{Bugli:2019rax}. We assume the radius of the iron-core, $r_0$, to lie somewhere in the range of $10^{3}-10^{4}$ Km (for practical calculations $r_0 \approx 2000$ Km \cite{Bugli:2019rax}) and model the magnetic field as, $B(r)=B_0$ for $r<r_0$, and $B(r)=B_0(r_0/r)^3$ for $r>r_0$ \cite{Bugli:2019rax}. The matter potentials for $V_e$, $V_{\mu}$, and $V_{\tau}$ are determined by the matter density $\rho$ and electron number fraction $Y_e$, which we obtain from a simulation of an 18 $M_\odot$ progenitor \cite{Fischer:2009af} at $t=4.37$ ms \cite{Tang:2020pkp}. See Fig.~\ref{matter-profile} for further details.

The spin-flip transition in the resonance region is dictated by the adiabaticity coefficient $\gamma_\alpha$, which is heavily influenced by $\mu_\nu B_0$. When taking into account the electron-fraction and density of matter in the SN [cf.~Fig.~\ref{matter-profile}], $\dot{V}_\alpha(r)$ varies as a function of $r$ only, while $\ddot{\phi}(r)$ can be arbitrary. For simplicity, we assume $|\ddot{\phi}| = 0$. This assumption is valid as long as the fluctuations around the average velocity $\dot{\phi}$ are smaller in scale than the spin-precession scale \cite{Krastev:1989ix}, which is approximately $\pi/\mu_\nu B_0 \sim \mathcal{O}(10)$ kilometers at the resonance layer for $\mu_\nu B_0 =\left(10^{-14} \mu_B \right) \left( 10^{12}\text{ Gauss} \right)= 10^{-2} \mu_B$ G.
We find the surface of the iron-core to be the most promising region\footnote{Note that the resonance criterion [cf. Eq.~(\ref{resonance})] cannot be achieved for  $\Dot{\phi}=0$ outside the region where neutrinos are produced, $r \gtrsim 100$ Km \cite{Janka:2017vlw}, and even when satisfied for $r \lesssim 100$ Km, it cannot be adiabatic unless considering unrealistically large magnetic moment and magnetic field combination.} as $\gamma_\alpha \gtrsim 1$  for values as small as  $\mu_\nu B_0 = 10^{-2} \mu_B$G, the smallest over the whole profile. 
This is due to the fact that $\rho$ decreases with $r^{-3}$ and $\Dot{\rho}$ decreases with $-r^{-4}$, so $|\Dot{V}_\alpha|$ reduces as the distance ($r$) from the SN centre increases. At $r=r_0$, the smallest $|\Dot{V}_\alpha|$ occurs when the highest possible field strength, $B(r_0)=B_0$, is present. 

At the surface of the iron-core, $Y_e \approx 0.5$ and $n_e \approx n_n$, see Fig.~\ref{matter-profile}. Thus, at this point, $V_e \approx 0.5 \sqrt{2} G_F n_e = -V_x$, where $x=\mu,\tau$. $V_e$ is a monotonically decreasing function in the range of $r \sim 10^{3}-10^{4}$ Km, such that $2$ $\text{m}^{-1} \gtrsim V_e \gtrsim 0.01$ $ \text{m}^{-1}$. Therefore, the rate of rotation of $B$ in the range $-0.01$ $\text{m}^{-1} \gtrsim \dot{\phi} \gtrsim -2$ $ \text{m}^{-1}$ could produce a resonance in the $\nu_e$ and $ \Bar{\nu}_e$ channels, see Eq.~(\ref{resonance}), while the rotation in the opposite direction $2$ $\text{m}^{-1} \gtrsim \dot{\phi} \gtrsim 0.01$ $ \text{m}^{-1} $ produces a resonance in the $\nu_{\mu,\tau}$ and $\Bar{\nu}_{\mu,\tau}$ channels. Note that the magnitude of $|\dot{\phi}| \sim 0.01 $ $ \text{m}^{-1} $ implies that for resonant conversion to occur, it is sufficient for $B$ to undergo approximately one revolution within a width of about one kilometer (which corresponds to the resonance layer). This observation aligns with the simulation presented in \cite{Bugli:2019rax}, where the magnetic field $B$ reverses its direction within a few kilometers around $r_0$.

\textbf{\emph{Analysis and results}.--}
\begin{figure*}
\centering
  \includegraphics[width=0.95\textwidth]{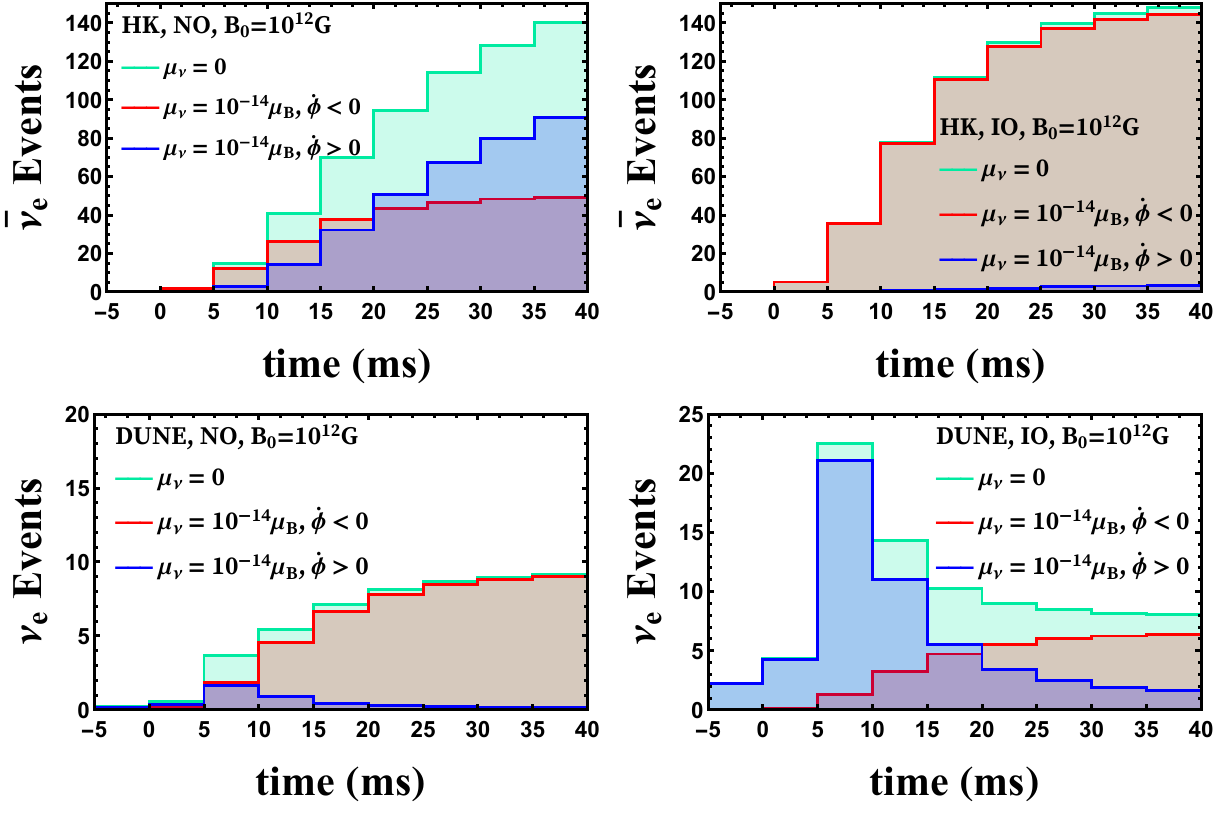}
  \caption{Expected number of $\bar{\nu}_e$-events at HK (upper) and $\nu_e$-events at DUNE (lower) for NO (left) and IO (right) in the time-window between $-5$ ms and $40$ ms corresponding to the SN neutronization-burst stage in which $0$ ms is the time of core-bounce. Results are shown for the standard scenario ($\mu_\nu=0$) and for $\mu_\nu=10^{-14} \mu_B$ and $B_0=10^{12}$ G for both cases~\ref{case1} ($\Dot{\phi}<0$) and \ref{case2} ($\Dot{\phi}>0$).}
\label{events-plots}
\end{figure*}
Before analyzing the effect of non-zero neutrino magnetic moments, we briefly describe the signal for the standard  scenario ($\mu_\nu=0$). SN neutrinos are produced at the core where $V_e \gg V_{x}$ and the electron-neutrino ($\nu_e$) becomes the heaviest of the three matter-eigenstates, either as $\nu_{3m}$ for normal mass ordering (NO) or as $\nu_{2m}$ for inverted mass ordering (IO); while the muon-neutrino ($\nu_\mu$) and tau-neutrino ($\nu_\tau$) are a combination of the two remaining eigenstates. The opposite is true for antineutrinos, where the electron antineutrino ($\bar{\nu}_e$) becomes the lightest of the three eigenstates in matter, either as $\Bar{\nu}_{1m}$ in NO or as $\Bar{\nu}_{3m}$  in IO, because $\bar{V}_{e} \ll \bar{V}_{x}$. During the subsequent evolution, neutrinos and antineutrinos might cross adiabatically the low ($L$) and high ($H$) Mikheyev-Smirnov-Wolfenstein (MSW) resonances \cite{Wolfenstein:1977ue,Mikheyev:1985zog,Mikheev:1987jp}. These happen when $|V_e-V_{x}|=|\Delta m^2_{n1}|\cos \theta_{1n}/2E$, where $n=2 (n=3)$ for $L (H)$ resonance. In NO, $L-$ and $H-$resonances occur exclusively in the neutrino channel, but in IO, $L-$resonance occurs for neutrinos and $H-$resonance for antineutrinos \cite{Dighe:1999bi}. Considering the best-fit values for the oscillation parameters \cite{deSalas:2020pgw}, $H-$resonance occurs at about $3 \times 10^4$ Km, whereas $L-$resonance occurs at approximately $2 \times 10^5$ Km from the centre of the SN. 
In NO, at the production region $\nu_e \approx \nu_{3m}$, so $\Phi_{i3} = \Phi_{ie}$. At vacuum, however, only a small component of $\nu_3$, $|U_{e3}|\approx 0.02$, remains as electron flavor. Therefore, only $0.02 \Phi_{ie}$ contributes to the final $\nu_e$-flux. Assuming the initial $\nu_\mu$-flux and $\nu_\tau$-flux to be equal, $\Phi_{i \mu}  = \Phi_{i \tau} = \Phi_{ix}$, the initial fluxes of the remaining matter-eigenstates $\nu_{1m}$ and $\nu_{2m}$, which are mixtures of $\nu_\mu$ and $\nu_\tau$, are also equal to $\Phi_{ix}$. Thus, their respective contributions to the final $\nu_e$-flux are $|U_{e1}|^2\Phi_{ix}$ and $|U_{e2}|^2\Phi_{ix}$. The total final $\nu_e$-flux in NO is then:
\begin{equation} \label{phi-e-no}
    \Phi_e^{NO}= |U_{e3}|^2\Phi_{ie}+(1-|U_{e3}|^2)\Phi_{ix} \approx  0.02 \Phi_{ie}+ 0.98\Phi_{ix},
\end{equation}
where we used the unitarity of the PMNS matrix to write $|U_{e1}|^2+|U_{e2}|^2=1-|U_{e3}|^2$. Similarly, for IO: $\Phi_{i2} = \Phi_{ie}$, while $\Phi_{i1}=\Phi_{i3}=\Phi_{ix}$, therefore:
\begin{equation} \label{phi-e-io}
    \Phi_e^{IO}= |U_{e2}|^2\Phi_{ie}+(1-|U_{e2}|^2)\Phi_{ix} \approx 0.3 \Phi_{ie}+ 0.7\Phi_{ix}.
\end{equation} 
For antineutrinos, at the production point, $\bar{\nu}_e \approx \Bar{\nu}_{1m}$ in NO and $\bar{\nu}_e \approx \Bar{\nu}_{3m}$ in IO. Then, we have:
\begin{equation} \label{phi-ebar-no}
    \Phi_{\bar{e}}^{NO}= |U_{e1}|^2\Phi_{i\bar{e}}+(1-|U_{e1}|^2)\Phi_{ix} \approx  0.7 \Phi_{i \bar{e}}+ 0.3 \Phi_{ix},
\end{equation}
\begin{equation} \label{phi-ebar-io}
    \Phi_{\bar{e}}^{IO}= |U_{e3}|^2\Phi_{i\bar{e}}+(1-|U_{e3}|^2)\Phi_{ix} \approx  0.02 \Phi_{i \bar{e}}+ 0.98\Phi_{ix}.
\end{equation}
\begin{table}[tb!] 
\centering 
\resizebox{0.48\textwidth}{!}{%
\begin{tabular}{|c|llll|} 
\hline
\hline
\multirow{3}{*}{\textbf{Experiments}} &
  \multicolumn{4}{l|}{\hspace{0.5cm}\textbf{Discovery reach for $\mu_\nu$ (in $\mu_B$)}} \\ \cline{2-5} 
 &
  \multicolumn{2}{c|}{\textbf{$\dot{\phi}<0$}} &
  \multicolumn{2}{c|}{\textbf{$\dot{\phi}>0$}}  \\ \cline{2-5} 
 &
  \multicolumn{1}{c|}{\textbf{NO}} &
  \multicolumn{1}{c|}{\textbf{IO}} &
  \multicolumn{1}{c|}{\textbf{NO}} &
  \multicolumn{1}{c|}{\textbf{IO}} \\ \hline
\textbf{HK} &
  \multicolumn{1}{l|}{$8 \times 10^{-15}$} &
  \multicolumn{1}{l|}{$\hspace{0.5cm}-$} &
  \multicolumn{1}{l|}{$8 \times 10^{-15}$} &
  \multicolumn{1}{l|}{$4 \times 10^{-15}$}  \\
   \hline
\textbf{DUNE} &
  \multicolumn{1}{l|}{$\hspace{0.5cm}-$} &
  \multicolumn{1}{l|}{$7.2 \times 10^{-15}$} &
  \multicolumn{1}{l|}{$1 \times 10^{-14}$} &
  \multicolumn{1}{l|}{$8.2 \times 10^{-15}$}  \\
  \hline
  \textbf{DUNE+HK} &
  \multicolumn{1}{l|}{$7.5 \times 10^{-15}$} &
  \multicolumn{1}{l|}{$7.2 \times 10^{-15}$} &
  \multicolumn{1}{l|}{$7 \times 10^{-15}$} &
  \multicolumn{1}{l|}{$4 \times 10^{-15}$}  \\
  \hline
  \hline 
\end{tabular}
}
\caption{Experimental sensitivities on $\mu_\nu$ for different benchmark scenarios and $B_0 = 10^{12}$ G.}
\label{summary_NMM}
\end{table}
Now we analyze the effect of the new resonance due to twisting magnetic fields at the surface of the SN iron-core. This resonance can happen at $r \sim r_0$, before the $L$ and $H$ resonances take place, and can have profound consequences to the fluxes, see Eqs.~(\ref{phi-e-no}-\ref{phi-ebar-io}), coming out of the collapsing star during the neutronization-burst phase. The reason is that a sizable fraction of active neutrinos could be converted to their right-handed counterparts, effectively decreasing the initial fluxes before they reach the $L$ and $H$ resonances, \textit{i.e.}, 
\begin{equation}
    \Phi_{i \alpha} \rightarrow e^{-\frac{\pi}{2}\gamma_\alpha} \Phi_{i \alpha},
\end{equation}
for the specific flavor $\alpha$ which has the resonance condition in Eq.~(\ref{resonance}) satisfied. The  Landau-Zener factor $e^{-\frac{\pi}{2}\gamma_\alpha}$ is the ``flip" probability that a transition happens between the states $\nu_{\alpha L} \leftrightarrow \nu'_{ R}$ at the resonance point.
In what follows, we analyze the limiting case of total adiabaticity, where $\gamma_\alpha > 1$. Indeed, Fig.~\ref{matter-profile} shows that for $\mu_\nu B_0=\left(10^{-14} \mu_B \right) \left( 10^{12}\text{G} \right)= 10^{-2}\mu_B\text{G}$, $\gamma_\alpha \approx 2$ and $e^{-\frac{\pi}{2}\gamma_\alpha} \approx 0.04$ in specific locations resulting in dramatic modifications to the expected neutronization-fluxes. 

We obtain the initial neutrino-fluxes from a spherically-symmetric $15 M_\odot$ progenitor simulation \cite{garching} and consider the time-interval between $-5$ ms and $40$ ms as corresponding to the neutronization-burst phase. We separate the detected signal in two periods: $-5$ ms $<t < 20$ ms that encompasses the neutronization-peak \cite{deGouvea:2019goq,Jana:2022tsa} and has initial fluxes that are roughly related by  $\Phi_{ie} \approx 10 \Phi_{ix}$ and $\Phi_{ix} \approx 10 \Phi_{i\Bar{e}}$, and $20$ ms $< t < 40$ ms which has initial fluxes $\Phi_{ie} \approx \Phi_{i\bar{e}} \approx \Phi_{ix}$. Here, we compute the percentage reduction for the neutronization-burst fluxes relative to the expectation described by Eqs.~(\ref{phi-e-no}-\ref{phi-ebar-io}) if the spin-flip resonance is crossed adiabatically:
\begin{enumerate}
    \item \label{case1} For $-0.01$ $\text{m}^{-1} \gtrsim \dot{\phi} \gtrsim -2$ $ \text{m}^{-1}$, the resonance would lie in the $\nu_e$ and $\Bar{\nu}_{e}$ channels. We can account for the effect by making $\Phi_{ie} \rightarrow 0$ in (\ref{phi-e-no}) and (\ref{phi-e-io}) while $\Phi_{i\Bar{e}} \rightarrow 0$ in (\ref{phi-ebar-no}) and (\ref{phi-ebar-io}):
        \begin{itemize}
            \item $t < 20$ ms: reduction of $17 \%$ for $\Phi_e^{NO}$, $81 \%$ for $\Phi_e^{IO}$, $19 \%$ for $\Phi_{\bar{e}}^{NO}$ and $0 \%$ for $\Phi_{\bar{e}}^{IO}$.

            \item $t > 20$ ms: reduction of $2 \%$ for $\Phi_e^{NO}$, $30 \%$ for $\Phi_e^{IO}$, $70 \%$ for $\Phi_{\bar{e}}^{NO}$ and $2 \%$ for $\Phi_{\bar{e}}^{IO}$.

        \end{itemize}

    \item \label{case2} For $2$ $\text{m}^{-1} \gtrsim \dot{\phi} \gtrsim 0.01$ $ \text{m}^{-1}$ there could be a resonance in the $\nu_{\mu,\tau}$ and $\bar{\nu}_{\mu,\tau}$ channels. The consequence is that $\Phi_{ix} \rightarrow 0$ in all Eqs. from (\ref{phi-e-no}) to (\ref{phi-ebar-io}):
        \begin{itemize}
            \item $t < 20$ ms: reduction of $83 \%$ for $\Phi_e^{NO}$, $19 \%$ for $\Phi_e^{IO}$, $81 \%$ for $\Phi_{\bar{e}}^{NO}$ and $100 \%$ for $\Phi_{\bar{e}}^{IO}$.

            \item $t > 20$ ms: reduction of $98 \%$ for $\Phi_e^{NO}$, $70 \%$ for $\Phi_e^{IO}$, $30 \%$ for $\Phi_{\bar{e}}^{NO}$ and $98 \%$ for $\Phi_{\bar{e}}^{IO}$.
            
        \end{itemize}
\end{enumerate}
These numbers indicate that twisting magnetic fields can impact the time-variation of the neutronization-burst signal. In particular, Case-1 optimizes the impact of the new resonance for Inverted Ordering (IO), resulting in up to an 81$\%$ reduction in the $\nu_e$ neutronization-peak, which occurs before 20 ms and is significant in the standard case for IO, see the difference between the green and red lines in the panel for DUNE, IO in Fig.~\ref{events-plots}. However, Case-2 has the highest potential for demonstrating the effects of twisting magnetic fields since it can cause a strong suppression (greater than 70$\%$) of neutrino-fluxes in almost every detection channel and time-window. 

Fig.~\ref{events-plots} depicts cases~\ref{case1} ($\dot{\phi}=-0.9 \hspace{0.1cm} \text{m}^{-1}$) and \ref{case2} ($\dot{\phi}=+0.9 \hspace{0.1cm} \text{m}^{-1}$) in future experiments like DUNE \cite{DUNE:2020zfm}, a $40$ Kt of liquid argon time-projection chamber, which mainly detects $\nu_e$ via $\nu_e+{ }^{40} \mathrm{Ar} \rightarrow{ }^{40} \mathrm{~K}^*+e^{-}$, and HK \cite{Hyper-Kamiokande:2021frf}, a $374$ Kt water-Cherenkov detector which detects mainly $\bar{\nu}_e$ via inverse-beta decay, $\bar{\nu}_e+p \rightarrow e^{+}+n$. The number of detected events of $\nu_\alpha$ per energy is
\begin{equation}
    \frac{d N_{\nu_\alpha}}{d E_r}=\frac{N_{\mathrm{t}}}{4 \pi R^2} \int d E_t \Phi_{\alpha}\left(E_t\right) \sigma_\alpha\left(E_t\right) W\left(E_r, E_t\right)
\end{equation}
where $N_t$ denotes number of target particles in the detector, $R=10$ Kpc is the distance between Earth and the galactic-SN, $\Phi_\alpha$ and $\sigma_\alpha$ represent $\nu_\alpha$-flux and interaction cross-section, and $W$ is energy-resolution function. For technical details, see Ref.~\cite{Jana:2022tsa}.
To study the sensitivity of DUNE and HK to a non-zero $\mu_\nu$ in presence of resonant spin-precession, we use the $\chi^2$-estimator
\begin{equation}
    \chi^2=\min _{\xi} \sum_{i=1}^n 2\left[(1+\xi) F_i-D_i+D_i \ln \left(\frac{D_i}{(1+\xi) F_i}\right)\right],
\end{equation} 
with $F_i$ and $D_i$ the number of events in the $i$-th time-bin for finite and null values of $\mu_\nu$, respectively. We perform a shape-only analysis where the normalization parameter $\xi$ varies in the range $[-1,1000]$ and adjust the normalization of the test hypothesis to the true hypothesis. In this way, the analysis concentrates on the time-variation of the signal. Discovery reaches for DUNE and HK, as well as the combined analysis of both experiments, are summarized in Table~\ref{summary_NMM}. The sensitivity of the combination of DUNE and HK can reach $\mu_\nu$ of order $4-7.5 \times 10^{-15}\mu_B$ at $90 \%$ C.L. for $B_0=10^{12}$ G \footnote{However, larger magnetic fields, up to $10^{16}$ G \cite{Palenzuela:2021gdo}, are possible in neutron star mergers, and the interplay of matter effects in the merger ejecta and twisting magnetic fields could impact the perceived initial flavor-ratios coming from these sources.}.

Interstellar magnetic fields only influence the overall flux-normalization \cite{Kopp:2022cug} and not its temporal variation. Hence, if supernova-flux is measured in the future and the flux-deficit is seen to be different in various time-bins, this will occur owing to the above-mentioned resonances caused by the magnetic field configuration's twisting structure. While the intriguing effect could appear in the observed (anti)neutrino spectra at HK and DUNE experiments, the reality is expected to be significantly more intricate, necessitating future research. Even with assumed knowledge of neutrinos as Dirac fermions and known mass ordering, uncontrollable factors, including other BSM effects, may influence the magnetic moment effect.

{\textbf {\textit {Final remarks.--}}} We have shown that if the magnetic field inside the supernova has a twisting structure, then the rotation of the magnetic field along the neutrino trajectory can induce a resonant spin-conversion, which will affect predictions for the event rates when detecting supernova neutrinos in future neutrino experiments such as DUNE and HK. If neutrinos are Dirac particles possessing large magnetic moments, this resonance effect will present the optimal avenue towards unravelling the scenario at hand.

\vspace{0.2in}
{\textbf {\textit {Acknowledgments.--}}}We thank Alexei Y. Smirnov  for useful discussions. 

\bibliographystyle{utcaps_mod}
\bibliography{reference}
\end{document}